\documentclass[epj,final]{svjour}

\usepackage{latexsym}
\usepackage{url}
\usepackage{amsfonts}
\usepackage{amsmath, amssymb}
\RequirePackage{graphicx}

\begin{document}
\title{$H_0$ tension: clue to common nature of dark sector?}
\author{V.G. Gurzadyan\inst{1,2}, 
A. Stepanian\inst{1}
}                     
%
%
\institute{Center for Cosmology and Astrophysics, Alikhanian National Laboratory and Yerevan State University, Yerevan, Armenia \and
SIA, Sapienza Universita di Roma, Rome, Italy}
\date{Received: date / Revised version: date}
%

\abstract{
The recently sharpened $H_0$ tension is argued not to be a result of data calibration or other systematic but an indication for the common nature of dark matter and dark energy. This conclusion is devised within modified weak-field General Relativity where the accelerated expansion of the Universe and the dynamics of galaxy groups and clusters are described by the same parameter, the cosmological constant. The common nature of the dark sector hence will result in intrinsic discrepancy/tension between the local and global determinations of values of the Hubble constant.   
} 
\PACS{
      {98.80.-k}{Cosmology}   
     } 
%
\maketitle

\section{Introduction}

Recent measurements \cite{Riess} increase the existing tension between the Hubble constant determinations from Planck satellite data \cite{Pl} and lower redshift observations;  the earlier studies and various approaches for resolving the tension are discussed in \cite{Riess}.

We will consider the $H_0$ tension within the approach of weak-field modified General Relativity (GR) which enabled the common description of the dark matter and dark energy by means of the same value of the cosmological constant \cite{GS1,G,GS3}. That approach is based on the Newton's theorem on the equivalency of the gravity of the sphere and of a point situated in its center and provides a natural way for the weak-field modification of GR, so that dark energy is described by the Friedmann-Lemaitre-Robertson-Walker (FLRW) equations while the dark matter in galaxy groups and clusters is described by the weak-field GR.

It is a principal fact that by now both the strong field GR has been tested by the discovery of gravitational waves, while the weak-field  effects such as at the frame-dragging are traced by measurements of laser ranging satellites \cite{Ciu}. The weak-field modifications we are discussing below are by now far from being tested at satellite measurements and therefore the dynamical features of the local universe including of the galactic dark halos \cite{Ge}, galaxy groups \cite{HB,GS3}, can serve as unique probes for such weak-field modifications of GR. Among other modified gravity tests are the accurate measurements of gravitational lenses \cite{GS2}, along with the effects in the Solar system \cite{KEK} or traced from large scale matter distribution \cite{E}.

Thus, we show that if the cosmological constant $\Lambda$ describes both the accelerated expansion and dark matter at galaxy cluster scales, then it will lead to the intrinsic discrepancy in the global and local values of the Hubble constant.

\section{Newton's theorem and $\Lambda$}

In \cite{GS1} it is shown that the weak-field GR can involve the cosmological constant $\Lambda$, so that the metric tensor components have the form
\begin {equation} \label {mod}
g_{00} = 1 - \frac{2 G m}{r c^2} - \frac{\Lambda r^2}{3};\,\,\, g_{rr} = (1 - \frac{2 G m}{r c^2} - \frac {\Lambda r^2}{3})^{-1}.
\end {equation} 
This follows from the consideration of the general function for the force satisfying Newton's theorem on the identity of sphere's gravity and that of a point situated in its center and crucially, then shell's internal gravity is no more force-free \cite{G1}. Namely, the most general form of the function for the gravitational force which satisfies that theorem is
\begin{equation}\label{force form}
F(r)= C_1r^{-2} + C_2 r,
\end{equation}   
where $C_1$ and $C_2$ are constants of integration; for derivation and discussion see \cite{G1,G}. The first term in Eq.(2) corresponds to the ordinary Newtonian law, and once the modified Newtonian law (for the potential) is taken as weak-field GR, one has Eq.(1), where the second constant $C_2$ corresponds to $\Lambda$ (up to a numerical coefficient and $c^2$) \cite{GS1,G}. Namely, the second constant $\Lambda$, on the one hand, acts as the cosmological constant in the cosmological solutions of Einstein equations, on the other hand, enters in the low-energy limit of GR which hence is attributed to the Hamiltonian dynamics of galaxy groups and clusters \cite{GS3}, instead of commonly used Newtonian potential. 

Within isometry group representation the Lorentz group O(1,3) acts as stabilizer subgroup of isometry group of 4D maximally symmetric Lorentzian geometries  and depending on the sign of $\Lambda$ (+,-, 0) one has the non-relativistic limits \cite{GS1} 

\begin{equation}\label{nrel}
\begin{aligned}
 & \Lambda>0: O(1,4) \to (O(3) \times O(1,1)) \ltimes R^6, \\  
 & \Lambda=0: IO(1,3) \to (O(3) \times R) \ltimes R^6, \\
 & \Lambda<0: O(2,3) \to (O(3) \times O(2)) \ltimes R^6.
\end{aligned}
\end{equation}
The O(3) is the stabilizer group for the spatial geometry since for all three cases the spatial algebra is Euclidean  
\begin{equation}
E(3)=R{}^{3} \rtimes O(3).
\end{equation} 
Thus, the Newton's theorem in the language of group theory can be formulated as each point of spatial geometry admitting the O(3) symmetry.

An important consequence of Eq.(2) is that the linear term (related to $C_2$ constant) can produce a non-zero force inside the shell. This is a unique feature since the pure Newtonian gravity according to Gauss' law cannot influence anything inside the shell. Furthermore, this mathematical feature of Eq.(2) can be considered as agreeing with the observational indications that the properties of galactic disks are determined by halos, see \cite{G}.

\section{Local and global Hubble flows with $\Lambda$}

The Hubble-Lemaitre law as one of established pillars of modern cosmology is characterized by the Hubble constant $H_0$ which
can be derived by various ways depending on the observational dataset. Namely, the Planck satellite provided the data on Cosmic Microwave Background (CMB) which within the $\Lambda$CDM model led to the following {\it global} value  $H_0= 67,66 \pm 0.42 km s^{-1} Mpc^{-1}$, as well as $\Lambda=1.11\, 10^{-52}m^{-2}$ \cite{P}. The recent analysis of Cepheid variables in Large Magellanic Cloud (LMC) by Hubble
Space Telescope (HST) \cite{Riess} led to the {\it local} value $H = 74.03 \pm 1.42 km s^{-1} Mpc^{-1}$. This discrepancy between the global and local values of the Hubble constant is the above mentioned tension.

Our Universe is considered to be described by FLRW metric
\begin{equation}\label{FLRW}
ds^2 = -c^2 dt^2 + a^2(t) (\frac{1}{1-kr^2}dr^2+r^2d\Omega^2),
\end{equation} 
where depending on the sign of sectional curvature  $k$, the spatial geometry can be  spherical $k=1$, Euclidean $k=0$ or hyperbolic $k=-1$. Consequently, the 00-component of Einstein equations for this metric is written as
\begin{equation}
H^2 =-\frac{k^2 c^2}{a^2(t)}+ \frac{\Lambda c^2}{3}+\frac{8\pi G\rho}{3}, 
\end{equation}
where $H=\dot{a}(t)/a(t)$ is the Hubble constant. 

Here an important point is the following. The Hubble-Lemaitre law originally was established for a sample of nearby galaxies,  which are members of the Local Group.  For them the empirical Hubble-Lemaitre law seemed to confirm the FLRW equations, however, later it became clear that not only the galaxies have their peculiar velocities but the Local Group itself is gravitationally bounded to a larger configuration, see \cite{Rai}. In other words, that law was observed at scales for which it should not be observed. Nevertheless, in spite of this apparent contradiction the local flow has been confirmed by observations: the detailed analysis of the nearby galaxy surveys reveal the local Hubble flow with $H_{loc}=78 \pm 2 km s^{-1} Mpc^{-1}$ \cite{Kar}.

We will now show that considering Eq.(1) as the weak-field limit of GR, it is possible to solve this tension.
Namely, the global Hubble flow will be described by the cosmological constant of FLRW metric, while the local flow by the weak-field GR given by Eq.(1). 

So, we are not allowed to use FLRW metric in local scales since the Local Supercluster galaxies do not move by FLRW geodesics. On the other hand due to attractive nature of pure Newtonian gravity one cannot produce a repulsive force to cause the local Hubble flow. However, if we consider the  additional linear term of Eq.(2) the $\Lambda$-term can cause a repulsive acceleration as
\begin{equation}
a = -\frac{GM}{r^2} +\frac{\Lambda c^2 r}{3}.
\end{equation}
It is simple to find out the distance at which the acceleration of ordinary Newtonian
term becomes subdominant with respect to the second term. In Table 1 the values for such
distances are listed for different mass scales. For objects less massive than the Local Group (LG), 
that {\it critical distance} is located outside the object's boundary, which means that it
cannot be observed. For LG, the critical distance is around 1.4 Mpc. Here, it is worth to mention that, since we have used Eq.(1) according to Newton's theorem, this distance can be considered as the radius of a sphere which the whole mass of LG is concentrated at its center. Thus, we conclude that, for those objects located outside this radius we will be able to observe an outward acceleration. These results obtained based on Newton's theorem are in agreement with other analysis \cite{Ch}.

\begin{table}
\caption{}\label{tab1}
\centering
\begin{tabular}{ |p{2.4cm}||p{2.1cm}|p{1.8cm}| }
\hline
\multicolumn{3}{|c|}{Critical distance for different objects} \\
\hline
Central Object& Mass (Kg)&Radius (m)\\
\hline
Earth &5.97 $\times 10^{24}$ & 4.92 $\times 10^{16}$  \\
\hline
Sun & 2 $\times 10^{30}=M_{\odot}$ & 3.42 $\times 10^{18}$ \\
\hline
Sgr A${}^{*}$  &4.3 $\times 10^6 M_{\odot}$&5.56 $\times 10^{20}$ \\
\hline
Milky Way &1.5 $\times 10^{12} M_{\odot}$ &3.91 $\times 10^{22}$ \\
\hline
Local Group  &2 $\times 10^{12} M_{\odot}$ &4.31 $\times 10^{22}$ \\
\hline
\end{tabular}
\end{table}
Meantime considering the weak-field limit according to Eq.(1), one can obtain the analogue of Eq.(6) for the non-relativistic case
\begin{equation}\label{WFH}
H^2 = \frac{\Lambda c^2}{3}+\frac{8\pi G\rho}{3}. 
\end{equation}
In spite of apparent similarity of Eq.(8) and Eq.(6), there is an important difference between them. Indeed,  in Eq.(8) $k=0$ and $\rho$ stands only for matter density (baryonic and non-baryonic), while $\rho$ in Eq.(6) includes the contribution also of radiation density; in this context see the comparative discussion on FLRW and McCrea-Milne cosmologies in \cite{CK}.
Thus, one can conclude that the $H$ observed by HST in local scales is not
the one obtained via Eq.(6) by considering the FLRW metric. It is a local effect which can be described by Eq.(8). However, before considering the weak-field limit equations for local flow, first let us take a look at the Eq.(1) itself.
According to principles of GR, the weak-field limit is defined when $\phi/c^2 \ll 1$, where $\phi$ is the weak-field potential. Now, by taking this into consideration, besides the Newtonian term a new limit is defined at large distances
\begin{equation}
 \frac{\Lambda r^2}{3} << 1, \,  r\simeq  1.46\, 10^{26} m = 5.33 Gpc.
 \end{equation}
Considering the fact that, the local Hubble flow is observed in few Mpc scales, we are allowed to use the Eq.(8) to describe that flow. By taking cosmological parameters \cite{P}, Eq.(6) confirms that the total matter density in our Universe is $\rho = 2.68\,  10^{-27} kg m^{-3}$. However,
by substituting $H = 74.03 \pm  1.42 km s^{-1} Mpc^{-1}$, the matter density which causes the
observed local Hubble flow will be $\rho_{loc} = 4.37 ^{+0.40}_{-0.39}\,  10^{-27} kg m^{-3}$. 

Now, in order to complete our justification we need to check the mean density of the local astrophysical structures. From hierarchical point of view the LG is located about 20 Mpc away from Virgo cluster\cite{Virgo}. The, Virgo cluster itself together with LG is in a larger Virgo supercluster \cite{VirgoSC}, which itself is the part of Laniakea supercluster \cite{Laniakea}. Considering the mass and their distances from LG, it is possible to find the distance where the density of these objects become exactly equal to $\rho_{loc}$. These results are exhibited in Table 2.

\begin{table}
\caption{}\label{tab2}
\centering
\begin{tabular}{ |p{2.7cm}||p{2.1cm}|p{2.1cm}| }
\hline
\multicolumn{3}{|c|}{Distances of objects where the density is $\rho_{loc}$} \\
\hline
 Object& Mass (Kg)&Distance from LG (Mpc)\\
\hline
Local Group &2 $\times 10^{12} M_{\odot}$ & $1.95\pm{0.06}$  \\
\hline
Virgo cluster & 1.2 $\times 10^{15}M_{\odot}$ & $3.45^{0.48}_{0.52} $ \\
\hline
Virgo supercluster  &1.48 $\times 10^{15} M_{\odot}$& $2.26^{0.51}_{0.56} $ \\
\hline
Laniakea &  $10^{17} M_{\odot}$ &$5.00^{2.07}_{2.29}$ \\
\hline
\end{tabular}
\end{table}
From these results it becomes clear that not only the error bars fully cover each other, but also the whole range of the local flow is covered by these values i.e. from 1.70 to 7.07 Mpc. Meantime, according to Eq.(7) the critical distance of Virgo supercluster from LG roughly is 7.27 Mpc, which means that the objects beyond that distance are gravitationally bounded to the supercluster. Considering the upper limit of Table 2 it turns out that there is no overlapping between the bounded objects and those who move away according to Eq.(8). Furthermore, these values exactly coincide with the density of Virgo cluster at distances in which Virgocentric flow changes to the FLRW linear Hubble-Lemaitre law \cite{Ch}.

Thus, the $H_0$ tension is not a calibration discrepancy but is a natural
consequence of presence of $\Lambda$ in GR as well as weak-field limit equations. While for
global value we have to consider the Eq.(6) as the immediate consequence of FLRW
metric and the cosmological parameters defined as
\begin{equation}
\Omega_k = -\frac{k^2c^2}{a^2(t)H^2}, \,\, \Omega_{\Lambda} =\frac{\Lambda c^2}{3H^2}, \,\, \Omega_m = \frac{8\pi G \rho}{3H^2}. 
\end{equation}
The local value of $H$ is obtained by weak-field limit equations and depends strictly on the
local density of matter distribution.

Note that, besides the above mentioned two evaluations
of $H$, other independent measurements also confirm this discrepancy. 
Among such measurements are those of the Dark Energy Survey (DES) Collaboration,
where the so-called inverse distance ladder method based on baryon acoustic oscillations
(BAO) is used \cite{DES}. Considering the BAO as a standard ruler in cosmology, it
turns out that its scale is roughly equal to 150 Mpc which clearly exceeds the
typical distance of our local structures (the Virgo cluster etc). Namely, the relevant SNe Ia are located at redshifts
$0.018 < z < 0.85$ \cite{DES}, which means that according to the Planck data \cite{P} such objects
are located at distances $80 Mpc < r < 3 Gpc$. Thus, by comparing these scales with the
typical distance to our local structures, one concludes that the measured $H$ for these observations should 
mainly be induced by cosmological parameters. This statement is justified by their
measured value $H = 67.77\pm 1.30 km s^{-1} Mpc^{-1}$.

Other measurements, again using BAO, are those of \cite{Ryan}, where like the DES
survey, the distances are $1.8Gpc < r < 6.2Gpc$ and yield $H = 67.6 ^{+ 0.91}_{-0.87} km s^{-1} Mpc^{-1}$. 

Thus, one can conclude that there are two different $H$s of two different scales,
{\it local} and {\it global} ones. Consequently, the measurement of these two quantities will depend on
scales attributed by the observations. Namely, for observations of
local scales it is expected to get the local $H$, while moving to cosmological scales i.e. 
beyond the Virgo cluster, the measurements should yield the  global $H$.

Note one more important point: although currently the numerical values of these two different $H$s are close to each other, 
their physical content is totally different. Namely, this semi-coincidence is due to the fact that, for the global case the density in
Eq.(6) is the current mean density in the Universe.  At earlier phases of the Universe the radiation density had a major contribution to the mean density
\begin{equation}
\Omega_{\rho}=\Omega_m+\Omega_r.
\end{equation}
Also, current observations \cite{P} indicate close to zero curvature of the Universe, k=0, and hence Eq.(6) will be similar to the weak-field equation. In other words, while for the local flow - no matter in which era - the contribution of matter density would have been the dominant one, for the global flow the contribution to the density in Eq.(6) was different for other cosmological eras where the radiation and $k$ were not negligible. 

Considering the FLRW metric's Hubble constant i.e. the global $H$ for different eras, one has 
\begin{equation}
H(t) = H_0 [\Omega_m a^{-3}(t) + \Omega_r a^{-4}(t) + \Omega_k a^{-2}(t) + \Omega_{\Lambda}]^{1/2},
\end{equation} 
where $H_0$ is the current value of the global Hubble constant. In this sense, the above statement
about the differences between $H$s will be also true as the Universe tends to de Sitter phase. In that case, all 
$\Omega$s except $\Omega_{\Lambda}$ will gradually tend to zero. But again, for the local measures one still will have the same non-zero matter density.

\section{Conclusions}

The $H_0$ tension can be not a result of data calibration/systematic but a genuine indication for the common
nature of the dark matter and dark energy. This conclusion is argued above based on the Newton's theorem and resulting
weak-field limit of General Relativity which includes the $\Lambda$ constant. Within that approach while the Friedmannian equations with the $\Lambda$ term are describing the accelerated Universe, the same $\Lambda$ is responsible for the dynamics of galaxy groups
and clusters.  Correspondingly, the {\it global} Hubble constant derived from the CMB and the {\it local} one devised from the galaxy surveys, including within the Local Supercluster, have to differ. 

Then, the long known so-called Local Hubble flow \cite{Kar} i.e. when the galaxies within the Local Supercluster are fitting the Hubble-Lemaître  law while the galaxies themselves are not moving via geodesics of FLRW metric, finds its natural explanation within the metric Eq.(1).  In other words, the local value of $H_0$ has to take into account the contribution of the cosmological constant (as entering the weak-field GR) in the kinematics of the galaxies along with the observed value of the mean density of matter.   

Accurate studies of the dynamics of galactic halos, groups and galaxy clusters, the gravitational lensing, can be decisive for further probing of described weak-field GR and the common nature of the dark sector.

\section{Acknowledgment}
We are thankful to the referee for valuable comments. AS is partially supported by the ICTP through AF-04.

\end{document}